\documentclass[aps,pra,epsfigure,twocolumn,showpacs]{revtex4}
\usepackage{amsmath}
\usepackage{amstext}
\usepackage{latexsym}
\usepackage{graphicx}
\usepackage{amsfonts}
\usepackage{epsfig}
\usepackage{color}

\begin{document}
\title{Noise resilient quantum interface based on QND interaction}
\author{Petr Marek}

\affiliation{Department of Optics, Palack\'{y} University, 17. listopadu 1192/12, CZ-771 46 Olomouc, Czech Republic}

\author{Radim Filip}

\affiliation{Department of Optics, Palack\'{y} University, 17. listopadu 50, 77200 Olomouc, Czech Republic}


\begin{abstract}
We propose a quantum interface protocol based on two quantum-non-demolition interactions (QND) arranged either in sequence or in parallel. Since the QND coupling arises naturally in interactions between light and a macroscopic ensemble of atoms, or between light and a micro-mechanical oscillator, the proposed interface is capable of transferring a state of light onto these matter systems. The transfer itself is perfect and deterministic for any quantum state, for arbitrarily small interaction strengths, and for arbitrarily large noise of the target system. It requires an all-optical pre-processing, requiring a coupling stronger than that between the light and the matter, and a displacement feed-forward correction of the matter system. We also suggest a probabilistic version of the interface, which eliminates the need for the feed-forward correction at a cost of reduced success rate. An application of the interface can be found in construction of a quantum memory, or in the state preparation for quantum sensing.
\end{abstract}

\maketitle
\section{Introduction}
Light is a natural carrier of information and since the advent of quantum information and metrology it solidified as a practical tool for quantum communication and quantum sensing. However, light is not sufficient on its own. For example, in a long distance quantum communication it has to be supplemented by quantum repeaters \cite{repeaters} based on, for example, atomic memory \cite{memory}. For the quantum sensing of magnetic fields, a cloud of atoms is used as a sensor, which is prepared and probed by the light beam \cite{sensing}. Light is also used as an efficient probe and resource for the preparation and monitoring of quantum vibrations of a mechanical oscillator \cite{mirrors}. It is therefore apparent that important elements of the quantum toolbox should be interface protocols, which allow for a flawless transfer from one physical system (light) to another (matter) \cite{interface}. One feature, this protocols should keep in regard, is the often limited capability of these matter systems to interact with light, be measured, and be controlled. The interfaces should therefore rely neither on pre-processing, nor on extensive post-processing of the matter system. Furthermore, the perfect interface should be achievable even for a weak coupling between the two physical systems and for a very noisy input state of the matter system.

Previous study of this problematic revealed that for almost any Gaussian coupling, with Hamiltonian quadratic in quadrature operators $x$ and $p$, a suitable pre-processing of the well-controllable system, measurement, and feed-forward can be used to implement a perfect state transfer without any regards for the initial state of the target system \cite{filip09}. This approach unfortunately does not work for the asymmetrical quantum-non-demolition (QND) coupling used to couple light to the macroscopic ensemble of atoms \cite{atoms}, or between light and macroscopic vibrations of the mechanical oscillator \cite{mirrors}. At its core, the QND interaction allows to transmit only a single quadrature from light to the matter system.  A method of working around this limiting factor and implementing a state transfer through a single QND interaction has been studied in Ref.~\cite{filip08}, but it was shown impossible for weak coupling strengths. Hence we will focus on the problem of building, in principle, a perfect interface for this kind of weak-strength QND coupling, for a moment neglecting other technical details of the light-matter interaction.

In this paper, we propose a deterministic universal quantum interface based on two sequential weak QND couplings between two light beams and a noisy matter system. The interface is based on all-optical (finite gain) pre-processing, a homodyne detection of light beams, and a conditional feed-forward correction displacing the matter system. By a construction we prove that the interface can perfectly transfer any (even unknown) quantum state to the initially noisy matter system, with no regards to strength of the QND couplings and the initial noise of the matter system. This interface method is then extended to a parallel QND coupling, represented by a sum of two QND couplings jointly probing the matter system.

We also present a probabilistic version of the interface, which is also able to transfer any state under the same conditions. In this scenario, there is no need for the feed-forward operation, a simplification, which is paid for by a reduced probability of success. We analyze the probabilistic scenario with regards to transfer of a single-photon state and demonstrate that weak coupling strengths and high initial noise can be compensated by a more severe post-selection.

\section{Deterministic interface}

The class of QND couplings between two harmonic oscillators is characterized by interaction Hamiltonian proportional to a product of two quadratures, one corresponding to a continuous-variable of light and the other to a continuous variable of matter. All the QND couplings (of the same strength) are theoretically equivalent, differing just by a simple local unitary transformation (phase-space rotation) of the relevant light and matter modes. Therefore, all the results presented below can be easily adapted for any kind of QND coupling. A single particular QND coupling can be characterized by the interaction Hamiltonian of the form $H = \chi x_A p_L$, where $\chi$ stands for the interaction strength, $x$ and $p$ denote the quadrature operators with $[x_{.},p_{.}] = i$, and the subscripts $L$ and $A$ mark the participating light mode and the matter mode (atomic ensemble mode or mechanical oscillator mode), respectively. In Heisenberg picture, the quadrature operators transform as
\begin{eqnarray}
    x_L^{out} = x_L^{in} + \kappa x_A^{in},~p_L^{out} = p_L^{in},\nonumber \\
    x_A^{out} = x_A^{in},~p_A^{out} = p_A^{in} -\kappa p_L^{in}.
\end{eqnarray}
The interaction transfers information about quadratures $x_A$ and $p_L$ to quadratures $x_L$ and $p_A$, while leaving the original quadratures undisturbed. This behavior is where the quantum-non-demolition coupling got its name. In following, we shall employ parameter $\kappa = \chi T$ (where $T$ is the effective lenght of the interaction) as the gain of the QND interaction. This simple unitary model ignores all the decoherence effects typical for the matter system, therefore it can only describe a coupling, which is very fast, faster than any relevant decoherence time. As a natural consequence, the range of values of $\kappa$ is very limited, typically corresponding to a weak coupling regime ($\kappa<1$).

As was mentioned previously, it is possible to conceive a noise free transfer using a single QND coupling, measurement and feed-forward, and a suitable squeezing of the initial state of light \cite{filip08}. However, there are several drawbacks to approach. The transfer can be noiseless, but it is still affected by loss and squeezing. The amount of loss depends on the strength of the coupling, with transmission parameter $\eta=\kappa^2/(1+\kappa^2)$, so unless the coupling gain is quite large, $(\kappa\gg 1)$, the loss will reduce the quality of the transfer. The squeezing can be compensated by another QND interaction with ancillary mode of light followed by a measurement and a feed-forward, but this incurs further loss and the overall transmission ends up only $\eta'=\kappa^2/(1+\kappa^2)^2$. All in all, there is no set of finite parameters for which lossless transfer is possible. Furthermore, the transfer is noiseless only when the mode of the matter system is prepaired in the ground state, which, despite the significant progress done by vacuum state preparation of the atomic ensemble or the mechanical oscillator, can still be a problematic task on its own.

\begin{figure}
\centerline{\psfig{figure=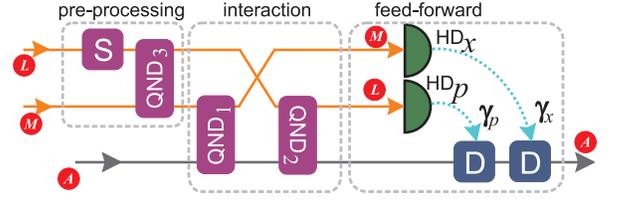,width=0.9\linewidth}}
\caption{(color online) Setup for the universal quantum interface based on the sequential QND couplings: $L$, $M$ stands for the light modes and $A$ denotes the matter mode. QND$_i$ - QND interaction with gain $\kappa_i$, HD$_q$, $q=X,P$ -  balanced homodyne detection measuring a quadrature $q$, $\gamma_q$ - gain in the feed-forward loop for quadrature $q$; D - displacement operation; S - optional squeezing operation. }\label{fig_2QND_setup}
\end{figure}

The need for the second QND coupling with an ancillary mode of light to compensate for the squeezing may be seen as an indication that a different scheme, which employs two sequential couplings with independent modes of light, may achieve better results. Clearly, these two couplings have to address both the complementary matter variables $x_A$ and $p_A$. Since the matter state is typically encoded up to a slowly rotating frame in the phase space, it is just a question of precise timing to correctly implement a sequence of the two couplings. Let us consider a system with two modes of light, $L$ and $M$ and one matter mode $A$. After the interaction mediated by the two sequential QND couplings characterized by interaction Hamiltonians $H_1=\chi_1 x_A p_M$, $H_2 = \chi_2 x_L p_A$ and gains $\kappa_1$, $\kappa_2$, the quadrature operators of the participating modes are transformed to
\begin{eqnarray}
  x'_M &=& x_M + \kappa_1 x_A, \nonumber \\
  p'_M &=& p_M, \nonumber \\
  x'_L &=& x_L, \nonumber \\
  p'_L &=& p_L - \kappa_2 p_A + \kappa_1 \kappa_2 p_M, \nonumber \\
  x'_A &=& x_A + \kappa_2 x_L , \nonumber \\
  p'_A &=& p_A - \kappa_1 p_M.
\end{eqnarray}
To complete the transfer we need to employ a pair of balanced homodyne detectors to measure values of the operators $x'_M$ and $p'_L$. This differs from the method of Ref.~\cite{filip09}, as there are no collective measurements required here, due to the specific nature of the QND interaction. The measured data are used in the feed-forward loop to drive the displacement of the mode $A$, $x_A^{out}= x'_A+\gamma_x x'_M$ and $p_A^{out}= p'_A + \gamma_p p'_L$, where $\gamma_x$ and $\gamma_p$ are the adjustable electronic gains. If these gains are chosen so $\gamma_x = -1/\kappa_1$ and $\gamma_p = 1/\kappa_2$, the influence of the initial state of the matter mode completely vanishes and its final state quadratures read
\begin{eqnarray}
x_A^{out}&=& \kappa_2 x_L - \frac{1}{\kappa_1} x_M , \nonumber \\
p_A^{out}&=& \frac{1}{\kappa_2} p_L.
\end{eqnarray}
As in \cite{filip09}, this dependance on collective quadratures of the initial modes of light can be removed by a suitable pre-processing. Namely, a squeezing of the mode $L$ with gain $g = \kappa_2$, followed by a QND coupling with interaction hamiltonian $H_3 = \chi_3 x_M p_L$ and gain $\kappa_3 = 1/(\kappa_1 \kappa_2)$, will achieve realization of a perfect transfer $x_A^{out} = x_L^{in}$ and $p_A^{out} = p_L^{in}$. Such the pre-processing can be built using an interferometric scheme based on either in-line squeezers or a measurement-induced QND (with the off-line squeezing), both of which are experimentally available \cite{QNDinline,QNDonline}. However, as of yet, the quality and possible gains of the pre-processing are limited by strengths of the in-line interactions or the available off-line squeezing.

A very interesting aspect of this protocol, a full depiction of which can be found in Fig.~\ref{fig_2QND_setup}, is that it can principally achieve a perfect transfer for an arbitrarily weak coupling between the light and the matter modes. Furthermore, although the demand on resources during the pre-processing phase can be substantial, as the gain $\kappa_3$ increases rapidly as $\kappa_1$ and $\kappa_2$ are getting smaller, the gains of both the QND transformation and the squeezing are always finite. Furthermore, the strength of interaction between light and matter is compensated for by a stronger all-optical coupling. Consequently, improvement in our ability to produce a more intensive coupling between optical modes would allow for a construction of better quantum interfaces between light and matter.

\begin{figure}
\centerline{\psfig{figure=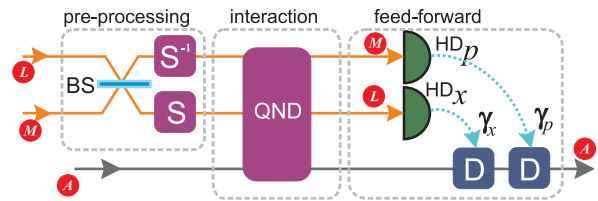,width=0.9\linewidth}}
\caption{(color online) Setup for the perfect state transfer. $L$, $M$ and $A$ denote the participating modes. HD$_q$ -  balanced homodyne detection measuring a quadrature $q$, $\gamma_q$ - gain in the feed-forward loop for quadrature $q$; S - squeezing operation; PS - phase shift operation, BS balanced beam splitter. }\label{fig_JQND_setup}
\end{figure}

A possible inconvenience of the presented method lies in its sequential nature. The light modes $L$ and $M$ are supposed to interact with the non-light mode $A$ one at the time and therefore the second one needs to be delayed while the first one interacts. This can be problematic if the light modes are defined over a substantially long time, which is a typical situation in experiments, where it allows to effectively increase the interaction strength.
A possible remedy lies in using a joint same-time QND interaction of the three modes, $M$, $L$ and $A$. A similar strategy was already proposed for a readout from the atomic memory \cite{Fiurasek06}. This coupling, consisting of the simultaneous addressing of the matter system by both the light systems and having the interaction Hamiltonian $H = \chi ( x_A p_L + p_A x_M)$, transforms the quadrature operators as:
\begin{eqnarray}
  x'_M &=& x_M, \nonumber \\
  p'_M &=& p_M - \kappa p_A + \frac{\kappa^2}{2} p_L, \nonumber \\
  x'_L &=& x_L + \kappa x_A + \frac{\kappa^2}{2}x_M, \nonumber \\
  p'_L &=& p_L , \nonumber \\
  x'_A &=& x_A + \kappa x_M , \nonumber \\
  p'_A &=& p_A - \kappa p_L,
\end{eqnarray}
where $\kappa$ is the gain of the joint QND coupling. After homodyne measurement of quadratures $x'_L$ and $x_M$, the state of the mode $A$ can be suitably displaced, $x_A^{out}= x'_A +\gamma_x x'_M$ and $p_A^{out}= p'_A +\gamma_p x'_L$, and if the gains are tuned properly, $\gamma_x = -\gamma_p = -1/\kappa$, the output state operators are
\begin{eqnarray}
  x_A^{out} &=& \frac{\kappa}{2} x_M  - \frac{1}{\kappa}x_L ,\nonumber \\
  p_A^{out} &=& \frac{1}{\kappa} p_M - \frac{\kappa}{2} p_L.
\end{eqnarray}
After a suitable pre-processing, which is depicted in Fig.~\ref{fig_JQND_setup}, and which consists of a balanced beam splitter and a pair of squeezing operations with gain $g = \sqrt{2}/\kappa$,
\begin{eqnarray}
  x_M &=& \frac{1}{\kappa}(x^{in}_M+x^{in}_L),\nonumber \\
  p_M &=& \frac{\kappa}{2}(p^{in}_M + p^{in}_L),\nonumber \\
  x_L &=& \frac{\kappa}{2}(x^{in}_M - x^{in}_L),\nonumber \\
  p_L &=& \frac{1}{\kappa}(p^{in}_M - p^{in}_L),
\end{eqnarray}
the quadratures of the input mode $L$ are perfectly transcribed onto the output mode $A$, $x_A^{out} = x_L^{in}$, $p_A^{out} = p_L^{in}$.

\section{Probabilistic interface}

The both already mentioned sequences of interactions, measurements, and conditional operations are capable of a deterministic transfer of an unknown quantum state from one physical system to another. However, for the state preparation it is often enough if the known quantum state can be transferred probabilistically, using a post-selection instead of the feed-forward. Specifically, instead of using the measured values of the operators $x'_M$ and $p'_L$ (or $x'_L$ and $p'_M$ for the simultaneous interaction) to drive the correcting displacement, we post-select the state of the target system if the detected values are zero and discard the state otherwise.  To accurately describe the projection nature of the measurement procedure we need to abandon the Heisenberg picture employed so far and look at the evolution of the actual physical states. Since the two scenarios are very similar, we shall focus only on the sequential QND interface illustrated in Fig.~\ref{fig_2QND_setup}.

First, let us check whether the probabilistic interface can be implemented in the limiting case of perfect transfer and pure states. These states can be described in coordinate representation by their wave functions, $|\psi\rangle = \int \psi(x)|x\rangle dx$, and any unitary evolution can be characterized by its action on the basis states. In our scenario, the global basis state is composed of three states $|x_L\rangle_L|x_M\rangle_M|x_A\rangle_A$ and our aim is to transform it to $|x_L\rangle_A$. In this representation, the separate parts of the evolution illustrated in Fig.~\ref{fig_2QND_setup} act as follows:
\begin{eqnarray}
\mbox{squeezing}:&\,\,\,& |x\rangle_L \rightarrow |x/g \rangle, \nonumber \\
\mbox{QND}_1:&\,\,\,&|y\rangle_M|z\rangle_A\rightarrow |y+\kappa_1 z\rangle_M|z\rangle_A,\nonumber\\
\mbox{QND}_2:&\,\,\,&|x\rangle_L|z\rangle_A\rightarrow |x\rangle_L|z+\kappa_2x\rangle_A,\nonumber\\
\mbox{QND}_3:&\,\,\,&|x\rangle_L|y\rangle_M\rightarrow |x+\kappa_3y\rangle_L|y\rangle_M.
\end{eqnarray}
After the squeezing $S$, an all-optical coupling QND$_3$, and sequential QND$_1$ and QND$_2$ interactions with the atomic system $A$, the states
$|x_L\rangle_L|x_M\rangle_M|x_A\rangle_A$ are transformed to
\begin{equation}
|x_L/g+\kappa_3\rangle_L|x_M+\kappa_1x_A\rangle_M|x_A+\kappa_2(x_L/g +\kappa_3 x_M)\rangle_A.
\end{equation}
After the projective measurements, which are represented by $_L\langle p = 0|_M\langle y=0|$, where $|p=0\rangle_L \propto \int_{-\infty}^{\infty}|x\rangle_L dx$, are applied to the optical modes $L$ and $M$, the resulting state is proportional (up to a normalization) to   \begin{equation}
\int_{-\infty}^{\infty}\hspace{-0.4cm}\delta(x-x_L/g-\kappa_3x_M)dx |-(\kappa_1^{-1}-\kappa_2\kappa_3)x_M+\kappa_2x_L /g\rangle_A.
\end{equation}
Now, the integral is equal to unity, and if we have chosen a suitable preprocessing with $g = \kappa_2$ and $\kappa_3 = (\kappa_1\kappa_2)^{-1}$ the final basis state is simply $|x_L\rangle_A$. As a result we have an atomic system corresponding perfectly to a wave function of the initial light system. The state of the optical mode $L$ has been perfectly transferred to the atomic mode $A$ without any regards for the initial state of any of the three participating modes.

The projective measurements can be approached by homodyne detectors followed by a post-selection if the measured values fell into a narrow interval around the desired values $p_L = 0$ and $x_M = 0$. However, the ideal situation, when the width of the interval approaches zero, also leads to a zero probability of success for the scheme. To realistically analyze the influence of the non-zero post-selection interval on the performance of the method and the probability of success, we will employ the formalism of Wigner functions \cite{wigner}. Wigner function of a single mode state is a real function of a pair of real variables $x$ and $p$, which directly correspond to the quadrature operators used before. The complete state of the three initial systems is represented by a joint Wigner function
\begin{equation}
    W_{in}(\xi) = W_L(x_L,p_L)W_M(x_M,p_M)W_A(x_A,p_A),
\end{equation}
where the $\xi$ is a shorthand notation of the vector of the variables, $\xi = (x_L,p_L,x_M,p_M,x_A,p_A)^T$. The subscripts $L,M,$ and $A$ again denote the three participating systems. The unitary evolution of the complete system can be expressed by a symplectic matrix $U$, which transforms the variables as $\xi' = U\xi$. Note that the same matrix can be used to implement the evolution for a vector of quadrature operators. The sequence of interactions depicted in Fig.~\ref{fig_2QND_setup} with $\kappa_3 = -1/\kappa_1\kappa_2$ is represented by matrix
\begin{equation}
    U = \left(\begin{array}{cccccc}
                0 & 0 & \kappa_1^{-1} & 0 & -1 & 0 \\
                0 & -\kappa_2^{-1} & 0 & 0 & 0 & -1 \\
                \kappa_1\kappa_2 & 0 & 1 & 0 & -\kappa_1 & 0 \\
                0 & (\kappa_1\kappa_2)^{-1} & 0 & 1 & 0 & \kappa_1^{-1} \\
                -\kappa_2 & 0 & 0 & 0 & 1 & 0 \\
                0 & 0 & 0 & \kappa_1 & 0 & 1
              \end{array}
              \right).
\end{equation}
After the transformation, the projective measurements are implemented as integration of the measured variables over interval $[-Q,Q]$, leading to Wigner function
\begin{equation}\label{Wig_out}
    W_{out}(x_A,p_A) = \frac{1}{PS}\int_{-\infty}^{\infty}\hspace{-0.4cm} dp_M \int_{-\infty}^{\infty}\hspace{-0.4cm} dx_L \int_{-Q}^{Q}\hspace{-0.4cm}dx_M \int_{-Q}^{Q}\hspace{-0.4cm} dp_L W_{in}(U \xi),
\end{equation}
where $PS$ is a normalization factor which also stands for the probability of success. A general evaluation of the integral is not an easy task. Fortunately, under a realistic assumption that the modes $M$ and $A$ are initially in thermal states with Wigner functions
\begin{eqnarray}\label{Wig_thermal}
  W_M(x,p) &=& \frac{\exp(-(x^2+p^2)/2 V_M)}{2\pi V_M} \nonumber \\
  W_A(x,p) &=& \frac{\exp(-(x^2+p^2)/2 V_A)}{2\pi V_A},
\end{eqnarray}
the limit $Q\rightarrow 0$ is indeed $W_{out}(x_A,p_A) = W_L(x_A,p_A)$, as predicted by the simple method.

To discuss the convergence of the method with regards to the post-selection interval $Q$ we need to consider a particular quantum state. The chosen state should be quite vulnerable to loss and noise, so we can clearly see their influence. This translates to a nonclassical non-Gaussian state, a clear specimen of which is the single-photon state with its trademark negativity in the Wigner function. The Wigner function itself is
\begin{equation}\label{Wig_sp}
    W_L(x,p) = \frac{\exp(-x^2-p^2)}{\pi} (2 x^2 + 2p^2 -1),
\end{equation}
and, under the assumption (\ref{Wig_thermal}), the final state can be analytically obtained from (\ref{Wig_out}). To evaluate the quality of the transfer we can calculate fidelity
\begin{equation}
    F = 2\pi \int_{-\infty}^{\infty}\hspace{-0.4cm} dx \int_{-\infty}^{\infty}\hspace{-0.4cm} dp W_{out}(x,p)W_L(x,p),
\end{equation}
to gain a measurable overlap between the ideal and the existing state. We can also explicitly try to look at transfer of the non-classical property --
the highly negative value of the Wigner function at the point of origin. For our purpose, we can define the negativity of the Wigner function as $N = \min[ W(x,p)]$.
\begin{figure}
\centerline{\psfig{figure=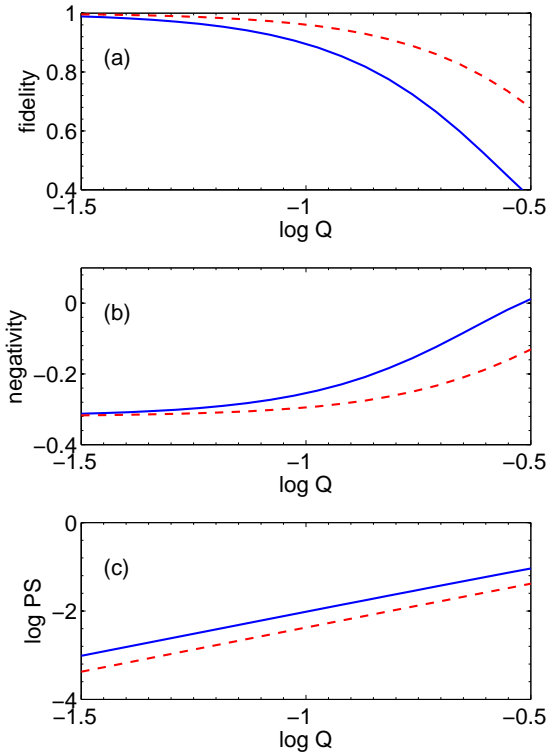,width=0.9\linewidth}}
\caption{(Color online) The fidelity \textbf{(a)}, the negativity of the Wigner function \textbf{(b)}, and the probability of success \textbf{(c)} for a transferred single-photon state as functions of the post-selection interval half-width $Q$. The parameters were $\kappa_1 = 0.3$, $\kappa_2 = 0.3$, $V_M = 0.5$, $V_A = 5$ for the blue solid line and $\kappa_1 = 0.5$, $\kappa_2 = 0.5$, $V_M = 0.5$, $V_A = 5$ for the red dashed line. }\label{fig_Qdependance}
\end{figure}

\begin{figure}
\centerline{\psfig{figure=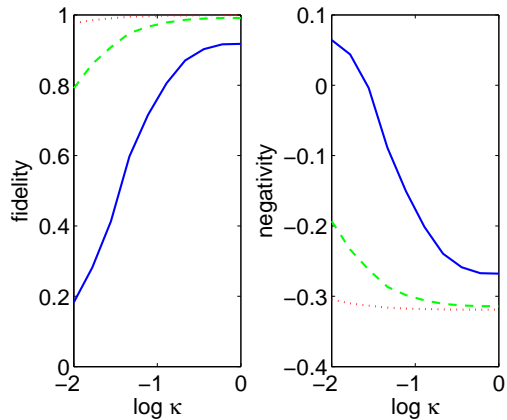,width=0.9\linewidth}}
\caption{(Color online) Values of the fidelity and the negativity of the Wigner function of a transferred
single-photon state relative to the coupling parameter $\kappa = \kappa_1
= \kappa_2$. The three curves correspond to different probabilities of
succes. Blue solid line: $PS = 10^{-2}$, green dashed line: $PS =
10^{-3}$, red dotted line: $PS = 10^{-4}$. Other parameters were: $V_M =
0.5$, $V_A = 5$.}\label{fig_kappadependance}
\end{figure}

\begin{figure}
\centerline{\psfig{figure=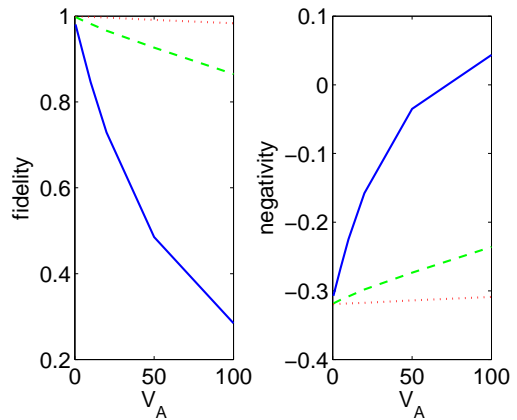,width=0.9\linewidth}}
\caption{(Color online) Values of the fidelity and the negativity of the Wigner function a transferred single
photon state relative to the level of initial noise in the atomic system
as characterized by $V_A$. The three curves correspond to different
probabilities of succes. Blue solid line: $PS = 10^{-2}$, green dashed
line: $PS = 10^{-3}$, red dotted line: $PS = 10^{-4}$. Other parameters
were: $V_M = 0.5$, $\kappa_1 = \kappa_2  = 0.5$.}\label{fig_VAdependance}
\end{figure}

These indicators, together with the probability of success, are shown in Fig.~\ref{fig_Qdependance} relative to the logarithm of the threshold value $Q$. By comparing the three panels we see that for a success probability of $PS \approx 0.01$ we can achieve fidelity $F \approx 0.9$ for both considered values of the coupling parameters. This may not seem much, but observe that the transferred state is still highly nonclassical, as witnessed by the negativity $N\approx -0.3$, which is only slightly worse than the ideal negativity of $N = - 1/\pi$. Further analysis revealed that the initial states of the $M$ and $A$ modes, as well as the QND coupling constants $\kappa_1$ and $\kappa_2$, are only relevant as far as the probability of success is concerned.
Fixing the success rate, the fidelity and the negativity are depicted in Figs.~\ref{fig_kappadependance},\ref{fig_VAdependance} as functions of coupling strength $\kappa=\kappa_1=\kappa_2$ and the initial noise of the matter system $V_A$. Both the characteristics are robust against low coupling and large noise, if the success probability is sufficiently low. This clearly demostrates a basic feasibility of the probabilistic transfer for the state preparation of highly non-classical states of atomic ensembles or micro-mechanical oscillators.

\section{Conclusion}
We have proposed a quantum interface capable of transferring a quantum state from one quantum system to another, for example, from light to a continuous-variable matter system, such as a collective spin mode of an atomic cloud or a vibrational mode of a mechanical oscillator. The main building blocks of the interface are QND interactions, which naturally arise as a coupling between modes of light and the aforementioned matter systems. The main benefit of the proposed scheme lies in its deterministic nature and its flawless performance - unit fidelity of transfer can be, in principle, achieved for arbitrary states of all participating modes and for arbitrarily small values of the QND coupling parameters. Furthermore, the resources required for the perfect transfer are always finite.

The interface can be also implemented probabilistically, thus eliminating the necessity of the feed-forward correction performed on the matter system. This is especially interesting for the state preparation of non-classical states of atoms or mechanical oscillators, where non-classical states of light are used as a resource. For a comparison, a single QND interaction, employed by cite{filip08}, allows to transmit a single-photon state perfectly, but a superposition of coherent states is transmitted with a reduction of the amplitude proportional to the strength of the QND coupling. The presently proposed method does not suffer from this -- the state of light is perfectly transferred to the atomic or mechanical system, without any feed-forward correction required.

Finally, this paper demonstrates a principal theoretical capability of constructing a perfect interface with help of QND couplings. Actual experimental considerations strongly depend on the employed physical system and will be addressed in a separate publication.

\medskip
\noindent {\bf Acknowledgments}
This research has been supported by projects  MSM 6198959213, LC06007 and Czech-Japan Project ME10156 (MIQIP) of the Czech Ministry
of Education. We also acknowledge grant 202/08/0224 and P205/10/P319 of GA CR and and EU grant FP7 212008 - COMPAS.

\end{document}